# Visualisation of sulphur on single fibre level for pulping industry

**B. Norlin**[a,1]**, S. An**[a]**, T. Granfeldt**[b]**, D. Krapohl**[a]**, B. Lai**[c]**, H. Rahman**[a]**, F. Zeeshan**[a] **and P. Engstrand**[a]

[a] *Mid Sweden University,*
  *Holmgatan 10, SE-851 70 Sundsvall, Sweden*
[b] *Valmet AB,*
  *Gustaf Gidlöfs väg 4, SE-851 79 Sundsvall, Sweden*
[c] *Argonne National Laboratory,*
  *9700 S. Cass. Avenue, Lemont, IL 60439, USA*

*E-mail:* borje.norlin@miun.se

ABSTRACT: In the pulp and paper industry, about 5 Mt/y chemithermomechanical pulp (CTMP) are produced globally from softwood chips for production of carton board grades. For tailor making CTMP for this purpose, wood chips are impregnated with aqueous sodium sulphite for sulphonation of the wood lignin. When lignin is sulphonated, the defibration of wood into pulp becomes more selective, resulting in enhanced pulp properties. On a microscopic fibre scale, however, one could strongly assume that the sulphonation of the wood structure is very uneven due to its macroscale size of wood chips. If this is the case and the sulphonation could be done significantly more evenly, the CTMP process could be more efficient and produce pulp even better suited for carton boards. Therefore, the present study aimed to develop a technique based on X-ray fluorescence microscopy imaging (µXRF) for quantifying the sulphur distribution on CTMP wood fibres.

Firstly, the feasibility of µXRF imaging for sulphur homogeneity measurements in wood fibres needs investigation. Therefore, clarification of which spatial and spectral resolution that allows visualization of sulphur impregnation into single wood fibres is needed. Measurements of single fibre imaging were carried out at the Argonne National Laboratory's Advanced Photon Source (APS) synchrotron facility. With a synchrotron beam using one micrometre scanning step, images of elemental mapping are acquired from CTMP samples diluted with non-sulphonated pulp under specified conditions. Since the measurements show significant differences between sulphonated and non-sulphonated fibres, and a significant peak concentration in the shell of the sulphonated fibres, the proposed technique is found to be feasible. The required spatial resolution of the µXRF imaging for an on-site CTMP sulphur homogeneity measurement setup is about 15 µm, and the homogeneity measured along the fibre shells is suggested to be used as the CTMP sulphonation measurement parameter.

---

[1] Corresponding author.

# Contents

## 1. Introduction

The impact of plastic packaging on the environment is becoming more and more concerning. Using wood fibre-based materials as a replacement could significantly reduce the problem since wood fibre-based materials decompose quickly. We anticipate High Yield Pulps (HYP) such as chemithermomechanical pulp (CTMP), to be an important component of sustainable packaging in the future. As estimated by the FAO Pulp paperboard grade annual report 2021, for carton board grades approximately 5.7 Mt of chemimechanical pulp (CMP/CTMP) was produced from softwood chips in 2020 [1]. The key unit operation of CTMP is to treat wood chips prior to defibration. The effectiveness of impregnation is crucial for pulp properties [2,3]. The efficiency of fibre separation in the chip refiner appears to be highly dependent on how uniformly the preheated chips are sulphonated. When impregnated, wood chips' inner parts accept a significantly lower amount of sulphonation than their outer parts [4]. However, it is unlikely to reach an even distribution of sulphite ($SO_3^{2-}$) ions in wood chips, resulting in inhomogeneous wood fibre properties [5]. Achieving better homogeneity of the sulphonation would reduce the sulphite dosage needed in the CTMP process. It is most likely that homogeneous sulphonation will reduce shive content, reducing energy consumption in the process [6]. Wood chips containing fibres with less or no sulphonation tend to fracture in the outer secondary cell wall, resulting in a carbohydrate-rich fibre surface with different bonding characteristics [7,8]. It is therefore necessary to distribute sulphur more evenly to improve product properties. It is possible to derive an elemental distribution map in a wide range of research disciplines by employing various analytical techniques. Among these are X-ray absorption spectroscopy [9], X-ray fluorescence spectroscopy (XRF) [10-14] and scanning electron microscopy and energy-dispersive X-ray spectroscopy (SEM-EDS) [15]. At Mid Sweden University, Sweden, XRF has already been used to verify the spectral resolution for simultaneous measurements of calcium (Ca) (3.7 keV) and copper (Cu) (8.0 keV) maps behind the paperboard coating [14].

However, the degree of unevenness is largely unknown at a micro scale level. At the micro scale level, it is still challenging to measure sulphonate distribution in wood chips and fibres. Partial defibration of woodchips before sulphonation, could produce a CTMP with more evenly distributed sulphur [16,17]. The hypothesis is that if sulphonation measurements should be useable for improving chip refinement, mapping of sulphur in fibres on a micro scale level is needed. By having this information, we would be able to gain a better understanding of sulphonation before defibration. Due to uneven wood chip surfaces and low fluorescence yields, our earlier ED-XRF in house measurements presented a challenge to detect light elements due to low fluorescence yield, low brilliance of the conventional X-ray source and intensity loss due to pinhole optics [18]. In these present measurements, the synchrotron radiation beam intensity at APS, USA, allowed excellent spatial resolution to detect the characteristic lines of light elements at 1 µm steps. APS's 2-ID-D beamline detected significant homogeneity differences in fibre sulphonation. The aim of these synchrotron measurements is to verify the hypothesis of uneven sulphonation at fibre level and to determine a suitable spatial resolution for on-site measurements. In this study, the objective is to propose a process to measure the homogeneity of the sulphur distribution in wood fibres.

## 2. Materials and methods

Measuring the degree of sulphonation on-site is an essential component of improving CTMP process system performance to achieve uniform properties at fibre level. ED-XRF (energy-dispersive X-ray fluorescence) imaging in an X-ray tube scanning setup is needed to achieve this goal. A setup using a sealed titanium box and helium atmosphere to improve spectral performance is considered in previous works [19]. It is, however, necessary to conduct synchrotron measurements to verify that this methodology is feasible.

### 2.1 Preparation of CTMP samples

To avoid influence from the uneven surfaces of wood chips, plain-surfaced paper samples or impregnated pulp samples are suggested for this experiment. The CTMP-712 reference Valmet pilot pulp was selected for investigation of sulphur distribution. A bleached softwood kraft pulp (BSWK) as SCA reference kraft K44 was used to dilute CTMP's sulphur density. Since the pulps had been thoroughly washed, it was assumed that unbleached kraft pulps (UBK) with lignin would not contain rest sulphur, which would result in completely sulphur-free BSWK. Low-grammage handsheets with 22 - 31 µm thickness made from a mix of four different proportions of CTMP and BSWK as seen in Table 1. In accordance with ISO 5269-2:2004, a low grammage handsheet of 10 g/m$^2$–14 g/m$^2$ was produced at SCA R&D Centre, Sundsvall, Sweden, using a conventional sheet former with a surface area of 0.021 m$^2$ [20].

**Table 1.** Samples ID used in measurements, CTMP/BSWK pulp ratio and imaged area.

| ID | CTMP/BSWK (%) | Area (mm$^2$) | ID | CTMP/BSWK (%) | Area (mm$^2$) |
|---|---|---|---|---|---|
| 31 | 50 / 50 | 0.0420 | 44 | 20 / 80 | 0.0605 |
| 33 | 50 / 50 | 0.0605 | 47 | 50 / 50 | 0.0557 |
| 36 | 40 / 60 | 0.0902 | 48 | 50 / 50 | 0.0515 |
| 37 | 40 / 60 | 0.1267 | 50 | 40 / 60 | 0.1232 |
| 43 | 20 / 80 | 0.0605 | | | |

## 2.2 APS Beamline

Sulphonation levels of a single fibre of CTMP paper sheets were examined using the APS (Advance Photon Source) USA's 2-ID-D beamline [21]. Sulphur distribution was mapped at a beam energy of 10.5 keV and a flux of $4\times10^9$ ph/s with a step size of 1 µm using a beam size of 0.3 µm $\times$ 0.3 µm. The energy resolution of the silicon drift detector is about 165 eV at 5.9 keV.

## 3. Results and discussion

The aim of this study is to verify what resolution of sulphur distribution is needed if a process relevant measured parameter should be extracted from the images. The challenge of measuring sulphur homogeneity is different depending on the size of order for the resolved part of the area. Since the diameter of the fibres in the images are in the range from 20 µm to 40 µm size, the different options for imaging are;

- Low spatial resolution not able to resolve fibres, but able to resolve distribution of sulphur on a global scale, revealing stirring variation in the pulp. This is expected if the spatial resolution is significantly larger than the fibre diameter.
- Spatial resolution able to resolve fibres, but not structures within the fibres. This is expected if the spatial resolution is approximately equal to the fibre diameter.
- Spatial resolution revealing structures within the fibres, requiring a resolution lower than the fibre diameter.

Based on these size considerations, some different methodologies to obtain a measurement parameter for the sulphur homogeneity are investigated, these are;

- Map the area of fibres containing sulphur and relate that area to the image area.
  - Advantage: Low spatial resolution, low requirement of image processing.
  - Disadvantage: Depending on the paper sheet, the density of the fibre varies greatly between images. No information is provided regarding single fibre levels.
- Use two measurement sets to get a reference of the area of the image containing fibres. Calcium is suggested, and an overexposed image is expected to reveal the full fibre area of the image. The sulphonated fibres can then be mapped against the calcium fibre area.
  - Advantage and disadvantage: Relatively low spatial resolution required, but two image sets with different energy selection.
- Gradients within the fibres to gain individual fibre sulphonation properties.
  - Advantage: Sulphonation on individual fibres retrieved
  - Disadvantage: High spatial resolution required.

According to our hypothesis we assume that sulphonation needs to be measured on individual fibre level.

### 3.1 Elemental distribution within the samples

The raw data from an XRF elemental mapping experiment is a fluorescence spectrum for each image pixel. It is possible to create images describing the density of an element's surface by selecting an energy range that corresponds to the element's characteristic energy. In Figure 1 such a set of images from the synchrotron measurement are exemplified. The first image reveals the sulphur distribution, and four individual fibres with high sulphur content are viewed together with one diagonal fibre with low sulphur content. The fourth calcium image reveals one more fibre in the top right part that is not visible in the sulphur image. Not all fibres are visible for all images of different elements. Calcium, copper, chlorine seem to be useful for viewing all fibres, while sulphur reveals the sulphonated parts of the samples. When fibres are crossing, the image shows the sum of the elemental content of one or more fibres. In Table 2, the maximum concentration for each element in each sample is listed. These values are calibrated at the APS synchrotron lab and give a size of order for the elemental content in the samples.

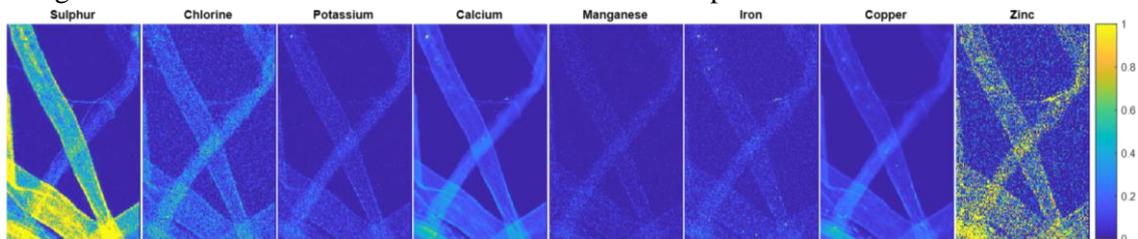

**Figure 1.** Synchrotron measurements for sample 31 in Table 1. Image map of the fluorescence energy of each element S, Cl, K, Ca, Mn, Fe, Cu and Zn. The colour bar describes the elemental concentration from zero to the Table 2 value for each element respectively (except underexposed Fe changed to 0.05 µg/cm$^2$).

**Table 2.** Highest measured surface concentration (µg/cm$^2$) for each element in each sample from Table 1. (Underlined value gives underexposed image due to single pixel outlier)

| Sample | 50 | 48 | 47 | 44 | 43 | 37 | 36 | 33 | 31 |
|---:|---|---|---|---|---|---|---|---|---|
| CTMP content | 40% | 50% | 50% | 20% | 20% | 40% | 40% | 50% | 50% |
| Sulphur | 4.29 | 4.30 | 2.41 | 4.03 | 2.73 | 3.27 | 1.86 | 2.00 | 2.40 |
| Chlorine | 0.371 | 2.66 | 0.418 | 1.53 | 0.548 | 1.63 | 0.384 | 0.441 | 0.385 |
| Potassium | 0.176 | 1.81 | <u>3.57</u> | 1.40 | 0.677 | 0.995 | 0.125 | 0.154 | 0.286 |
| Calcium | 4.76 | <u>53.0</u> | 30.8 | <u>120</u> | 8.08 | 15.6 | 16.8 | 7.09 | 5.95 |
| Manganese | 0.0387 | <u>0.679</u> | 0.0817 | <u>0.158</u> | 0.0729 | 0.0624 | <u>0.214</u> | 0.0199 | 0.0558 |
| Iron | <u>0.233</u> | 0.896 | 4.73 | 0.707 | <u>1.95</u> | 1.88 | <u>0.301</u> | 0.0737 | <u>0.755</u> |
| Copper | 4.09 | 2.68 | 1.58 | 3.81 | 5.02 | 0.919 | 1.41 | 1.42 | 0.945 |
| Zinc | 0.0339 | <u>0.289</u> | <u>3.97</u> | <u>0.602</u> | <u>0.829</u> | <u>1.64</u> | <u>0.267</u> | <u>0.105</u> | 0.0038 |

### 3.2 Sulphur homogeneity on samples

The sulphur distribution for all samples measured at the synchrotron are shown in Figure 2.

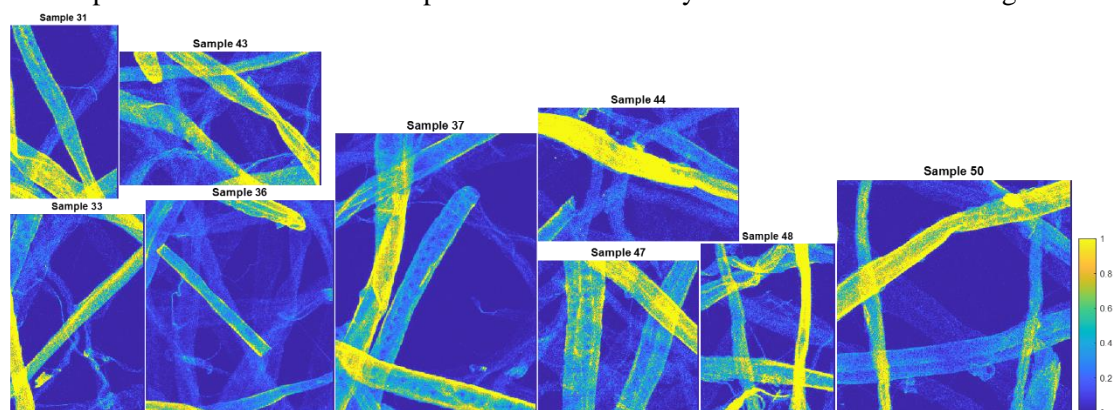

**Figure 2.** Synchrotron measurements of sulphur distribution for all samples in Table 1. The colour bar describes the sulphur concentration from zero to the Table 2 value for each sample respectively.

### 3.2.1 Manual fibre counting

The number of fibres in each image can be counted manually. However, due to the stacking of fibres in some areas of the images, this will provide low accuracy. A high fluctuation of the sul-phonated fibre ratio is seen in Table 3, and there is no significant correlation between the percentage of CTMP pulp and the sulphonated fibre ratio. Likewise, the sulphonated fibre area also shows high fluctuation.

**Table 3.** Estimated number of fibres in images, from manual counting. The top right and the top left part of Figure 1 and related images shows fibres containing sulphur or Calcium

| Sample | 50 | 48 | 47 | 44 | 43 | 37 | 36 | 33 | 31 |
|---|---|---|---|---|---|---|---|---|---|
| CTMP content | 40% | 50% | 50% | 20% | 20% | 40% | 40% | 50% | 50% |
| Sulphur fibres | 2 | 3 | 5 | 5 | 2 | 5 | 2 | 4 | 5 |
| Calcium fibres | 9 | 10 | 8 | 17 | 14 | 17 | 10 | 7 | 10 |
| S / Cl ratio | 0.22 | 0.30 | 0.63 | 0.29 | 0.14 | 0.29 | 0.2 | 0.57 | 0.50 |
| Image area (mm$^2$) | 0.042 | 0.061 | 0.090 | 0.127 | 0.061 | 0.061 | 0.058 | 0.052 | 0.123 |
| Fibres /cm$^2$ | 2.14 | 1.65 | 0.89 | 1.34 | 2.31 | 2.80 | 1.73 | 1.36 | 0.81 |
| S Fibres /cm$^2$ | 0.48 | 0.50 | 0.55 | 0.39 | 0.33 | 0.83 | 0.35 | 0.78 | 0.41 |

### 3.2.2 Sulphur sum and sulphur area calculation

The sum of sulphur content can be performed in two ways. One way is to average the content in all pixels, that will give a sulphur average value displayed in Table 4. The average is between 5% and 10% of the maximal sulphur values from Table 2. As for the manual fibre counting, the sulphur weight in Table 4 shows no significant correlation with the percentage of CTMP pulp. From these observations, measurements averaging sulphur over one sample seems to have limited usability.

**Table 4.** Sulphur weight average for the full area image of each sample.

| Sample | 50 | 48 | 47 | 44 | 43 | 37 | 36 | 33 | 31 |
|---|---|---|---|---|---|---|---|---|---|
| CTMP content | 40% | 50% | 50% | 20% | 20% | 40% | 40% | 50% | 50% |
| Sulphur average (µg/cm$^2$) | 0.231 | 0.369 | 0.319 | 0.329 | 0.297 | 0.225 | 0.133 | 0.195 | 0.261 |

Another methodology to measure the sulphur distribution is to use a sulphur concentration threshold and calculate the relative sulphonated area consisting of pixels above the concentration threshold, as seen in Figure 3. These area images can be interpreted as if the middle image is representing the sulphur distribution relatively well corresponding to the original image in Figure 2. The latter of these two images shows distinct imaging of the shell surrounding the fibres. The area calculated for all samples are presented in Table 5. As stated before for sulphur weight average, the sulphur area fraction also seems to have limited usability.

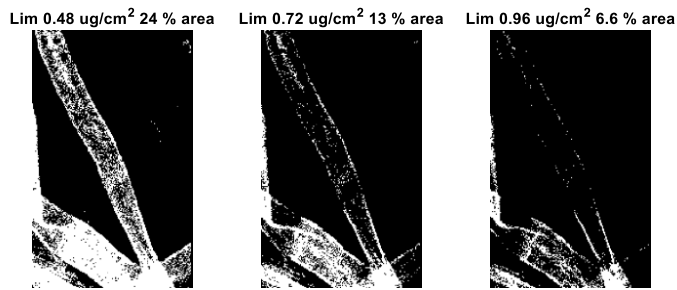

**Figure 3.** Binary sulphur area image for 3 different (Lim) thresholds concentrations from sample 31.

**Table 5.** Sulphur area threshold dependence, fraction of sample area in Table 1.

| Sample | Threshold | 50 | 48 | 47 | 44 | 43 | 37 | 36 | 33 | 31 |
|---|---|---|---|---|---|---|---|---|---|---|
| CTMP content | | 40% | 50% | 50% | 20% | 20% | 40% | 40% | 50% | 50% |
| Rel. Sulphur area | 0.48 | 0.286 | 0.401 | 0.273 | 0.263 | 0.368 | 0.294 | 0.131 | 0.219 | 0.245 |
| Rel. Sulphur area | 0.72 | 0.213 | 0.319 | 0.129 | 0.210 | 0.274 | 0.205 | 0.084 | 0.171 | 0.130 |
| Rel. Sulphur area | 0.96 | 0.161 | 0.242 | 0.058 | 0.186 | 0.194 | 0.138 | 0.054 | 0.134 | 0.069 |

### 3.2.3 Sulphur area measurement versus actual fibre area

To be able to relate the sulphonation to the actual fibre content in the fibres, the area occupied by fibres needs to be calculated. It is possible to use the calcium energy for showing all fibres in the samples, not only the fibres with high sulphonation. Figure 4 indicates that thresholds between 0.06 µg/cm$^2$ and 0.45 µg/cm$^2$ will reveal reasonable areas for the fibres.

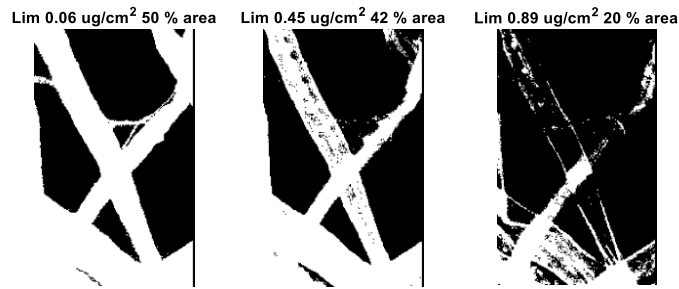

**Figure 4.** Binary calcium area image for 2 different (Lim) thresholds concentrations from sample 31.

**Table 6.** Calcium based fibre area vs threshold, fraction of sample area in Table 1.

| Sample | Threshold | 50 | 48 | 47 | 44 | 43 | 37 | 36 | 33 | 31 |
|---|---|---|---|---|---|---|---|---|---|---|
| CTMP content | | 40% | 50% | 50% | 20% | 20% | 40% | 40% | 50% | 50% |
| Rel. Calcium area | 0.06 | 0.628 | 0.690 | 0.875 | 0.818 | 0.897 | 0.574 | 0.810 | 0.535 | 0.500 |
| Rel. Calcium area | 0.45 | 0.427 | 0.526 | 0.719 | 0.584 | 0.725 | 0.443 | 0.533 | 0.384 | 0.421 |
| Rel. Calcium area | 0.89 | 0.267 | 0.412 | 0.472 | 0.430 | 0.562 | 0.250 | 0.294 | 0.235 | 0.197 |

When comparing the calcium-based fibre areas in Table 6 with the sulphur weight in Table 4, the fibre area adjusted Sulphur weight in Table 7 can be calculated. Still, no significant correlation between this area adjusted sulphur weight and CTMP content can be found. This indicates that a sulphonation measurement parameter must be from variations within single fibres.

**Table 7.** Sulphur (µg/cm$^2$) weight adjusted for actual fibre area.

| Sample | Threshold | 50 | 48 | 47 | 44 | 43 | 37 | 36 | 33 | 31 |
|---|---|---|---|---|---|---|---|---|---|---|
| CTMP content | | 40% | 50% | 50% | 20% | 20% | 40% | 40% | 50% | 50% |
| Sulphur in fibres | 0.06 | 0.368 | 0.535 | 0.365 | 0.402 | 0.311 | 0.392 | 0.164 | 0.364 | 0.522 |
| Sulphur average | 0.45 | 0.541 | 0.702 | 0.444 | 0.563 | 0.410 | 0.508 | 0.250 | 0.508 | 0.620 |

### 3.3 Microscopic sulphur distribution within single fibres

Visual observation of the sulphonation images reveals that most of the fibres have significantly higher concentration of sulphur attach to the fibre shell, while the fibre core only contains a moderate fibre concentration. This is the expected image projection of a circular tube with its main density in the outer shell, like a pipe geometry. However, a few fibres appear to have even radial distribution, like the two bottom left fibres on images for sample 31 and 33 in Figure 2. Possible explanations of the even distribution might be either a cracked fibre shell causing easy sulphonation of the core, or that the fibre is flattened so the image reveals the shell concentration, and no pipe projection is possible.

From a process perspective, our hypothesis evolved during this work, is that even sulphonation along the fibre shell is the most favourable parameter to extract as a process parameter. Such a parameter can be retrieved by measuring the variations along the outer parts of the fibre areas. The fibre periphery lines are extracted from the difference of eroded and dilated images. Fibre shell lines along 4 different fibres are manually selected, and the line width is chosen to be 4 pixels (4 µm) to achieve slight filtering of the shell sulphonation lines. Zoom in images and line scans of the selected fibres are seen in Figure 5 and in Table 8 data extracted from the figure is shown. The last three lines of Table 8 shows suggestions for sub-fibre sulphonation measurement parameters. The range is a measure of how even the sulphonation along the fibre shell. If this parameter is weighted using the average along the shell, low range due to low concentration (seen in Line 7 and Line 8) is corrected for. Since these parameters work along the fibre shell, a parameter comparing the shell sulphonation with the core sulphonation to get a radial distribution is also suggested. The radial parameter shows that Line 2 and Line 4 has weak sulphonation compared to the fibre core. Further studies are needed to evaluate the suggested parameters, since the CTMP fibres exemplified are equal in sulphonation processing.

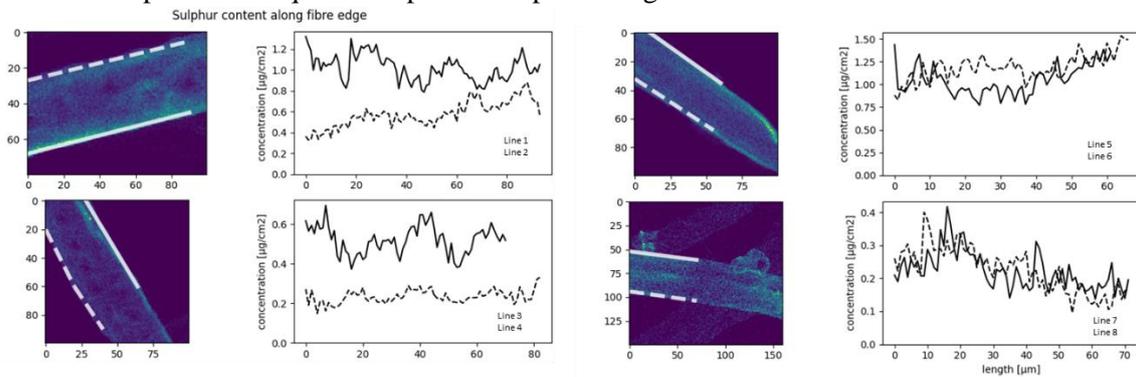

**Figure 5.** Sulphonation in the fibre shell along fibres. 4 µm thick shell measured.

**Table 8.** Sulphonation data extracted from fibre shell line scans.

|  | Line 1 | Line 2 | Line 3 | Line 4 | Line 5 | Line 6 | Line 7 | Line 8 |
|---|---|---|---|---|---|---|---|---|
| Avg. Conc. Fibre area | 0.338 | 0.338 | 0.133 | 0.133 | 0.401 | 0.401 | 0.0725 | 0.0725 |
| Avg. Conc. Fibre shell | 1.019 | 0.573 | 0.518 | 0.231 | 1.046 | 1.186 | 0.223 | 0.225 |
| Min. Conc. Fibre shell | 0.788 | 0.326 | 0.372 | 0.149 | 0.770 | 0.830 | 0.137 | 0.0956 |
| Max. Conc. Fibre shell | 1.321 | 0.881 | 0.693 | 0.288 | 1.437 | 1.535 | 0.418 | 0.400 |
| Shell half range | 0.267 | 0.278 | 0.161 | 0.070 | 0.334 | 0.353 | 0.141 | 0.152 |
| Shell half range / avg | 0.262 | 0.484 | 0.310 | 0.301 | 0.319 | 0.297 | 0.630 | 0.677 |
| Shell avg / Fibre avg | 3.01 | 1.69 | 3.89 | 1.74 | 2.61 | 2.96 | 3.08 | 3.10 |

**3.4 Spatial resolution needed for on-site homogeneity measurements**

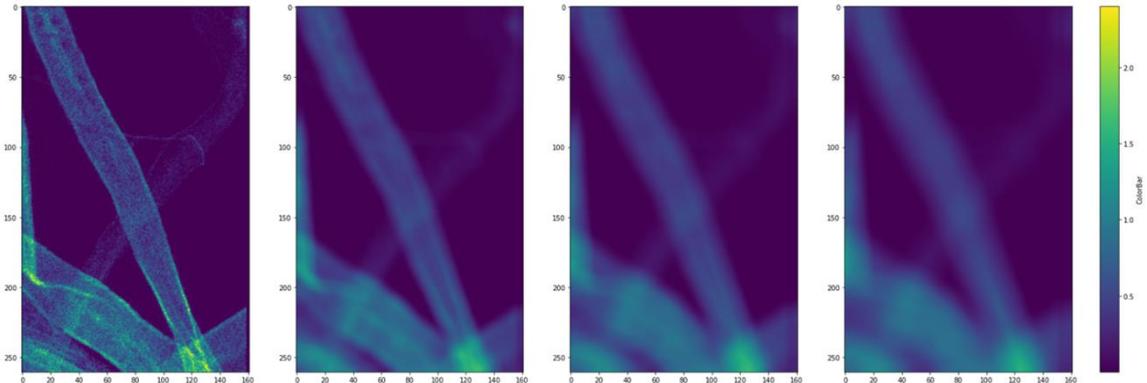

**Figure 6.** Resolution needed for sulphur imaging of sample 31; (a) Original synchrotron image with 1 µm step, (b) 10 µm spot size (c,) 15 µm spot size; (d) 20 µm spot size.

When moving from high resolution synchrotron measurements, to X-ray tube measurements, the spatial resolution of sulphur will deteriorate. Figure 6 shows how degenerating the spatial resolution blurs information about sulphur distribution inside the fibres, and eventually also the visualization of single fibres is lost. A system consisting of an X-ray tube equipped with polycapillary focusing optics must hence be capable of about 10 µm resolution [22]. Apart from spatial resolution, the spectral resolution also needs consideration [18].

## 4. Conclusion

The synchrotron images investigate the impregnation of sulphur into wood fibres for varying production parameters. Images of CTMP samples show a significant uneven distribution of sulphur between fibres from the perspective of the pulp process. Sulphur impregnation is also predominantly concentrated in the fibre shell at the individual fibre level. The resolution that still contains homogeneity information is extracted for further development of the methodology. The necessary spatial resolution is estimated between 10 µm and 15 µm for CTMP sulphur homogeneity inspection.

From a process perspective, our hypothesis evolved during this work, is that even sulphonation along the fibre shell is the most favourable parameter to extract as a process parameter. Such a parameter can be retrieved by measuring the variations along the outer parts of the fibre areas.

**Acknowledgments**

The authors acknowledge funding from Vinnova as project "*Renewable packaging materials - Impregnation depth measurements for pulping industry using a synchrotron.*", as well as discussions with representatives from the companies Billerud AB and Valmet AB. Synchrotron measurement was conducted in collaboration with the Advance Photon Source, a U.S. Department of Energy (DOE) Office of Science user facility operated for the DOE Office of Science by Argonne National Laboratory under Contract No. DE-AC02-06CH11357.